# Lattice-dynamics and in-plane antiferromagnetism in Mn$_x$Zn$_{1-x}$PS$_3$ across the entire composition range

Robert Oliva,[1] Esther Ritov,[2] Faris Horani,[2] Iñigo Etxebarria,[3] Adam K. Budniak,[2] Yaron Amouyal,[4] Efrat Lifshitz[2] and Mael Guennou[1]

[1]Department of Physics and Materials Science, University of Luxembourg, 41 rue du Brill, L-4422 Belvaux, Luxembourg

[2]Schulich Faculty of Chemistry, Solid State Institute, Russell Berrie Nanotechnology Institute, Helen Diller Quantum Center, Nancy and Stephen Grand Technion Energy Program, Technion−Israel Institute of Technology, Haifa 3200003, Israel

[3]Fisika Saila and EHU Quantum Center, Euskal Herriko Unibertsitatea UPV/EHU, Sarriena Auzoa z/g, 48940 Leioa, Basque Country, Spain

[4]Department of Materials Science and Engineering, Technion–Israel Institute of Technology, Haifa 3200003, Israel

**Abstract**

Alloyed Mn$_x$Zn$_{1-x}$PS$_3$ samples have been grown covering the whole compositional range and studied by means of Raman spectroscopy at temperatures covering from 4K up to 850K. Our results, supported by SQUID magnetic measurements, allowed, from one hand, to complete the magnetic phase diagram of Mn$_x$Zn$_{1-x}$PS$_3$ and establish $x{\geq}0.3$ as the composition at which the alloy retains antiferromagnetism and, from the other hand, to identify the Raman signatures indicative of a magnetic transition. The origin of these Raman signatures is discussed in terms of spin-phonon coupling resulting in the appearance of low- and high-frequency zone-folded phonon modes. For the alloy, an assignment of the 1$^{st}$ and 2$^{nd}$ order modes is provided with the aid of first-principle lattice-dynamical calculations. The compositional dependence of all phonon modes is described and the presence of zone-folded modes is shown to take place for both, the alloy and MnPS$_3$. Finally, a comparison of the Raman spectra of ZnPS$_3$ to other compounds of the transition-metal phosphorous trisulfide family allowed shows that low-frequency phonon peaks exhibit an abnormally large broadening. This is consistent with previous claims on the occurrence of a second-order Jahn-Teller effect that takes place for ZnPS$_3$ and Zn-rich Mn$_x$Zn$_{1-x}$PS$_3$.

## I. INTRODUCTION

The family of layered transition metal phosphorus trichalcogenides (TMPTs) gained renewed interest due to its unique magnetic properties together with excellent chemical and structural stability.[1] While early studies were devoted towards the understanding of the crystal structure, intercalation properties,[2] and their applicability as cathodes for lithium batteries,[3] it was later shown that TMPTs exhibit a plethora of types of two-dimensional magnetic ordering. This is interesting from both a fundamental and an applied perspective due to their potential for designing multifunctional materials and flexible electronics based on heterostructures and alloys.[4] Despite the high technological interest, TMPTs are scarcely investigated compared to other two-dimensional families such as transition metal dichalcogenides.

Transition metal phosphorous trisulfides with chemical formula $MPS_3$ (M typically being Mn, Fe, Co, Ni, Zn, Cd) crystallize into the monoclinic layered structure that corresponds to the C2/m space group. Single-layer $MPS_3$ consists of a cation metal ($M^{2+}$) arranged in a honeycomb array sandwiched between the chalcogen planes, while the adjacent layers are held by weak van der Waals S-S interlayer interactions.[5,6] Aside from robust chemical stability at ambient conditions, magnetic properties can be tuned via cation exchange by using divalent magnetic metal cations such as $Mn^{2+}$, $Fe^{2+}$, or $Ni^{2+}$, which exhibit short-range spin ordering at temperatures higher than the critical temperature. The magnetic properties of TMPTs are determined by the number of up-spins and their magnetic moment arrangement within a single layer. Spins mainly arrange antiferromagnetically with Néel,[5,7] stripy[8,9] or zigzag[10] disposition of 1st next neighbor spins (NNs). Importantly, super-exchange interactions with 2nd and 3rd NNs are not negligible and aid in controlling the magnetization stabilization resulting in Néel temperatures ($T_N$) in the range from 78 K, for MnPS$_3$, up to 155 K, for NiPS$_3$.[11]

Alloying allows to cover a wide band gap range further increasing the range of applicability of TMPTs, spanning from 1.6 eV, for magnetically-active NiPS$_3$, up to 3.4 eV, for diamagnetic ZnPS$_3$. The band gap in bulk compounds is either direct or quasi-direct and its nature depends on the degree of ionicity of the transition metal element. The impact of alloying magnetically-ordered MPS$_3$ into a diamagnetic matrix has been scarcely investigated. Early studies[12] evaluated the effect of alloying and temperature on the magnetic susceptibility for Fe$_x$Zn$_{1-x}$PS$_3$ (0< $x$ ≤1) and reported an increase of the Néel temperature upon increasing iron concentration. Dilution of spin 5/2 Mn$^{2+}$ ions on a diamagnetic host lattice of CdPS$_3$ showed that Mn(II) substitutionally replaces Cd and exhibits a high degree of covalency with NN sulfur ligands.[13] Later studies investigated the Mn$_x$Zn$_{1-x}$PS$_3$ alloy throughout the whole compositional range and confirmed that Zn cations randomly substitute in the crystal lattice.[14] Besides, the compositional effect on the $T_N$ was also evaluated for a few samples and the spin-flip field was investigated.[15,16] These works allowed to conclude that the 1st NN exchange is found to be independent of composition, while the critical concentration for long-range order was estimated to be 45% Mn. These effects were suggested to strongly impact the magnetic phase diagram so it is highly desirable to evaluate magnetic ordering along the whole compositional range.

Raman spectroscopy is a powerful tool to investigate not only the vibrational and structural properties of two-dimensional systems but also determine magnetic texture in 2D magnetic materials. Temperature-dependent Raman experiments on MPS$_3$ (M= Ni, Fe) and their alloyed compounds revealed rich Raman features below $T_N$ due to magnetic ordering. These effects were classified into three categories: *i)* Folding of the Brillouin zone (BZ) due to the presence of magnetic ordering, *ii)* spin-spin and spin-phonon interactions, and *iii)* interference of the single-phonon state with electronic transitions due to the spin splitting of the electronic band structure.[17–20]

In the present work we provide a comprehensive study on the effect of alloying magnetically-active MnPS$_3$ into a diamagnetic matrix (ZnPS$_3$) in order to shed new light on the magnetic, vibrational, and structural properties of alloyed TMPTs. Mn$_x$Zn$_{1-x}$PS$_3$ is particularly interesting from a fundamental perspective because, from one side MnPS$_3$ is a true 2D antiferromagnet (Heisenberg-type, with a spin angle of ≈8° from the out-of-plane direction[21]) that exhibits Néel-type magnetic ordering (i.e. spins are flipped between NNs)[10] at least down to the bilayer limit[22] and adjacent layers are coupled ferromagnetically[21] with proven magnon-spintronics capabilities.[23] While, on the other hand, ZnPS$_3$ is diamagnetic and might exhibit a distorted crystal lattice while maintaining a similar band gap and lattice parameters to those of MnPS$_3$. Note that previous experiments showed that a similar compound, CdPS$_3$, which belongs to the same group of Zn (i.e. IIb group), exhibits a distorted crystal lattice and a structural transition at T = 228 K,[24,25,26] so temperature-dependent measurements on ZnPS$_3$ are desirable. Finally, it is particularly interesting to research $Mn_xZn_{1-x}PS_3$ from a lattice-dynamical perspective because its cation average atomic number (from Z=25 for Mn up to Z=30 for Zn) covers those of the most relevant $MPS_3$ (i.e. $FePS_3$, $CoPS_3$ and $NiPS_3$), which could reveal valuable information with regard the relative impact of; *i)* the reduced atomic mass and *ii)* the electronic configuration on the Raman spectrum.

## II RESULTS AND DISCUSSION

*Lattice-dynamical calculations*

Bulk MnPS$_3$ and ZnPS$_3$ are layered compounds with ABC-type stacking. The corresponding space group is monoclinic, C2/m and the point group is C$_{2h}$. This structure gives rise to 15 Raman-active modes, from the irreducible representation, $\Gamma$ = 8A$_g$+6A$_u$+7B$_g$+9B$_u$, It can be seen that there are eight A$_g$ modes (here labeled from A$_{g1}$ to A$_{g8}$) which can be probed under parallel scattering configuration (e.g. $z(xx)\bar{z}$ in the conventional setting) and seven B$_g$ modes (here labeled from B$_{g1}$ to B$_{g7}$) which are active in cross-scattering configuration (e.g. $z(xy)\bar{z}$).

For ZnPS$_3$, the calculated frequencies and corresponding phonon dispersion curves (PDCs) are shown in Fig. 2 (calculated phonon frequencies of A$_g$ and B$_g$ modes are included as star and circle symbols, respectively). From the figure, very flat dispersion curves can be observed for the optical phononic branches (the chosen k-path is shown in Fig. 1-b with red lines and includes representative high-symmetry k-points), giving rise to a large phonon gap in the 330-540 cm$^{-1}$ region. This is in good agreement with previous lattice-dynamical calculations based on phenomenological models and reflects the distinct nature of the high-frequency modes (i.e. above 250 cm$^{-1}$), involving ion movements of the P$_2$S$_6$ octahedral cage, from that of the low-frequency modes (i.e. below 250 cm$^{-1}$), involving phonons with strong contributions of the heavy metal ions and phosphorus atoms.[2] However, while those calculations predicted a flat dispersion for the high-frequency modes, our first-principle calculation methods revealed that the frequency of the mode with the highest frequency, A$_{g8}$ significantly increases (by up to ≈35 cm$^{-1}$) at the border of the BZ. For the case of monolayers, calculations based on the DFT predicted a similar increase in the frequency of the highest-frequency mode in the M high-symmetry point for a slab, which is equivalent to the V point in the C2/m bulk structure.[27]

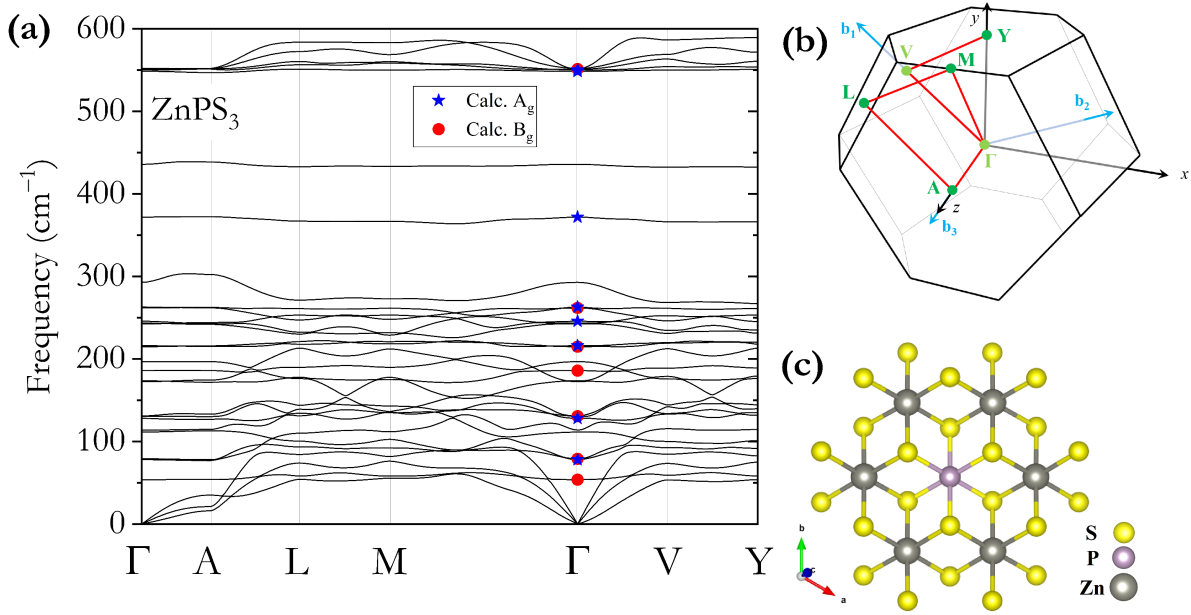

**Fig. 1. (a)** Calculated phonon dispersion curves of bulk ZnPS$_3$. Calculated frequencies of the A$_g$ and B$_g$ modes are shown as blue and red symbols, respectively. **(b)** First BZ with selected high-symmetry points together with the k-path used for the PDCs (red lines). **(c)** The top view of a ZnPS$_3$ slab is represented using the VESTA software.[28]

The phonon dispersion curves of MnPS$_3$ and frequencies of the A$_g$ and B$_g$ modes are shown in Fig. 2 (calculated frequencies of A$_g$ and B$_g$ modes are included as star and circle symbols, respectively). Similarly to ZnPS$_3$, a very flat phononic dispersion can be observed for most phononic branches and a frequency increase takes place for the highest modes at the border of the Brillouin zone. This is highly relevant for MnPS$_3$ since below the Néel temperature, zone folding phenomena induced by the elastic "Bragg" magnetic scattering from the spin superstructure is expected for phonon modes that either modulate the spin-orbit coupling or the exchange interaction.[29] The corresponding metal-related zone-folding phenomenon is schematically shown in Fig. 2-c for a slab. By observing the PDCs in the Γ-V direction for both, ZnPS$_3$ and MnPS$_3$, our calculations predict that zone-folding would give rise to the following new Raman features; *i*) a large number of modes in the 100-300 cm$^{-1}$ region, *ii*) a new Raman-active mode ≈33 cm$^{-1}$ above A$_{g8}$ and iii) a new Raman-active mode ~68 cm$^{-1}$ above A$_{g6}$.

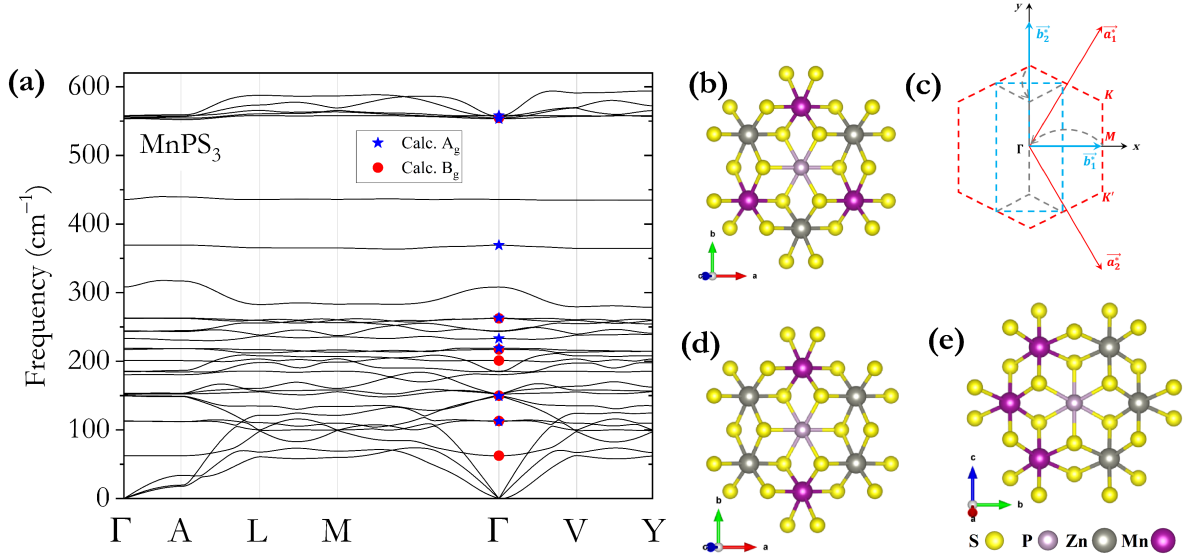

**Fig. 2. (a)** Calculated phonon dispersion curves of bulk MnPS$_3$. Calculated frequencies of the A$_g$ and B$_g$ modes are shown as blue and red symbols, respectively. **(b)** Top view of a MnZnP$_2$S$_6$ slab (note that this mixed crystal corresponds to a composition of Mn$_{0.50}$Zn$_{0.50}$PS$_3$) with an atomic arrangement corresponding to the crystallographic s.g. C2. Such arrangement results in a zone folding (panel **c**) of the BZ that is equivalent to antiferromagnetically-ordered MnPS$_3$. Two additional configurations of MnZnP$_2$S$_6$ with space groups P2/m and P2/c have been investigated and are plotted in panels **(d)** and **(e)**, respectively.

For the case of the alloy ($[x]$=50%) phonon frequencies at zone-center were calculated for the three structures shown in panels b, d and e of Fig. 2. The atomic arrangements presented in Fig. 2-b,d,e result in a decrease of symmetry from the C2/m structure of the compositional end-members, to P2/m, C2 and P2/c space groups, respectively. Consequently, the number of phonon-modes are increased at Γ. For the C2 structure, the phonon-folding from symmetry reduction results in $A_g$ and $A_u$ modes of C2/m becoming 16 A-symmetry modes and $B_{2g}$ and $B_{2u}$ become 14 E-symmetry modes, with a total of 27 Raman-active optical modes and 3 acoustic modes. The P2/c and P2/m structures exhibit a total of 60 and 58 modes, respectively (3 of which are acoustic) with $A_g$, $A_u$, $B_g$ and $B_u$ symmetries. Our calculations revealed that in-plane FM and AFM ordering exhibit energetic differences in the same order as different atomic arrangements (in the order of 10 of meV per 20 atoms), being the AFM systematically energetically favorable as well as the P2/c structure. The least favorable structure was P2/m with FM spin ordering (+67 meV for 20 atoms with respect to the P2/c-AFM structure). Most of their phonon frequencies nearly match the average between those calculated for the pure compounds at Γ and V (all zone-center phonon frequencies of the pure compounds as well as the three mixed arrangements are included in table SM1 for $A_g$ and $A_u$ symmetry equivalent modes and in table SM2 for $B_g$ and $B_u$ symmetry equivalent modes, calculated mode frequencies at V are shown in table SM3 of the SM for the pure compounds). Remarkably, out of the 14 optical $A_g$ and 16 $B_g$ calculated Raman-active modes of the P2/c structure, 18 modes exhibit nearly identical frequencies to those averaged for the compositional end members at Γ and V. Moreover, 2 modes correspond well with prominent expected folded-modes (i.e. a new mode 37.0 cm$^{-1}$ above $A_{g8}$ and a new Raman-active mode ~63.4 cm$^{-1}$ above $A_{g6}$). Hence, it could be argued that clustering at the atomic scale would enhance the Raman signal from prominent folded-modes, except for the folded mode 5.92 cm$^{-1}$ below the $A_{g7}$ mode which is only allowed for P2/m structure with first-neighbor atomic intercalation (Néel-type arrangement).

**Temperature-dependent Raman measurements of MnPS$_3$ and ZnPS$_3$**

Bulk MnPS$_3$ and ZnPS$_3$ are expected to give rise to 8 $A_g$ and 7 $B_g$ Raman-active modes. Since many of their phononic frequencies are nearly degenerate resulting in ten Raman peaks, these are labeled from $P_1$ up to $P_{10}$. Within this notation, these are shown for ZnPS$_3$ and MnPS$_3$ in Fig. 3-a) and 4-b), respectively, at temperatures from ambient temperature down to 4 K. The intensities of the spectra have been normalized to that of the most intense peak, $P_8$.

While the Raman spectra of MnPS$_3$ are relatively well understood both at room and low temperatures, the Raman spectra of ZnPS$_3$ have been comparatively scarcely investigated.[30,31] Strikingly, many Raman features of ZnPS$_3$ are strongly distinctive from those of MPS$_3$ with M = Mn, Ni, Co, Fe. The most notorious differences from MPS$_3$ compounds are that ZnPS$_3$ exhibits; *i)* a much larger broadening of the low-frequency modes (from $P_1$ to $P_7$) at ambient temperature, *ii)* a decrease in frequency from 20 cm$^{-1}$ and up to 60 cm$^{-1}$ for the $P_2$ and $P_3$ peaks when compared to other MPS$_3$ and *iii)* smaller frequency difference of the modes $P_6$-$P_7$ and $P_9$-$P_{10}$. This can be seen by comparing both panels of Fig. 3 as well as from Fig. SM2 of the SM. Such differences cannot be merely attributed to differences in the ionic mass or size of the metal since the atomic numbers of $M^ZPS_3$ compounds (Z = 25 to 28) are very similar to that of $ZnPS_3$ (Z=30).

The remarkable difference of the Raman spectra of $ZnPS_3$ in relation to any other TMPT can be accounted for by either of the following hypothesis; *i)* the crystal lattice of $ZnPS_3$ is not C2/m, or *ii)* the crystal lattice of $ZnPS_3$ is structurally disordered. Here we support the latter since previous XRD studies on nanocrystals found that $ZnPS_3$ is isomorph to compounds of the same family.[32,33] On the other hand, octahedrally-coordinated $d^{10}$ close-shell cations such as $Ag^+$, $Cu^+$, $Cd^{2+}$ or $Zn^{2+}$ typically exhibit coupling between the filled $d^{10}$ orbitals with energetically closely-lying empty s orbitals which decrease the energy barrier towards lower symmetry geometries and is experimentally measured as large thermal parameters and positional disorder (i.e. dynamic and static distortions, respectively).[34,35,36] Abnormally high atomic displacement parameters (ADPs) previously measured by single-crystal X-ray measurements allowed to confirm the presence of such second-order Jahn-Teller effect on $CdPS_3$ and $ZnPS_3$.[37,38] Moreover, a phase transition due to a re-arrangement of layer stacking from the low temperature trigonal R3 phase to the monoclinic C2/m phase have been shown take place for $CdPS_3$ at $T$ = 228 K,[24,25,26] but no equivalent studies have yet been performed for $ZnPS_3$. From our Raman spectra we conclude that the broadening of the low-frequency phonon modes, which mostly involve metallic ion displacements, is explained by the instability of Zn, in agreement with experimentally observed large ADPs.[37,38] As expected from anharmonic effects, the peak broadening increases up to 550 °C (high-temperature Raman spectra are shown in the SM Fig. SM6), at which temperature the sample starts to degrade, but no abrupt changes in broadening or peak frequencies are observed, thus indicating that no structural phase transition takes place for $ZnPS_3$. On the other hand, from Fig. 3-a) it can be seen that the broadening of the low-frequency peaks strongly decreases at lower temperatures and a strong increase of frequency takes place for P2 (i.e. around 8 cm$^{-1}$ which is one order of magnitude than typical shifts for higher-frequency peaks or peaks in $MnPS_3$). This evidences that our samples are highly crystalline, no structural transition takes place and that the dynamic disorder is corrected at low temperatures. These results are in perfect agreement with our DFT calculations, which confirmed that the stable phase for $ZnPS_3$ is C2/m at zero and room temperature since the potential well for Zn does not split when volumes are increased to that of $ZnPS_3$ at room temperature.

For $MnPS_3$, distinct Raman signatures abruptly show up right below the Néel temperature (as can be seen in Fig. 3-b, Fig. 4 and Fig. SM9-top of the SM), namely; *i)* the vanishing of P$_2$ around 117 cm$^{-1}$, *ii)* four new peaks around 194 cm$^{-1}$, 370 cm$^{-1}$, 565 cm$^{-1}$ and 605 cm$^{-1}$ and *iii)* many very weak Raman peaks in the range 125-300. Since these peaks are not visible right above the $T_N$, these cannot be attributed to first or second order Raman processes and must be inherently related to the magnetic transition. Here we attribute all these features to zone-folded phonon modes. This is a consequence of the doubling of the unit cell due to nearest-neighbor antiferromagnetic coupling (the so-called AFM-Néel type),[7] which results in the folding of the Raman phonon branches in the in-plane direction (as schematically shown in Fig. 2-c for the case of a 2D MnPS$_3$ slab). The phonon-folding phenomena have been already reported in *zigzag* antiferromagnetically ordered FePS$_3$[39] but for the case of MnPS$_3$, where first neighbors are alternatively arranged, the crystal symmetry is higher resulting in an effective folding of the BZ in the Γ-V direction.

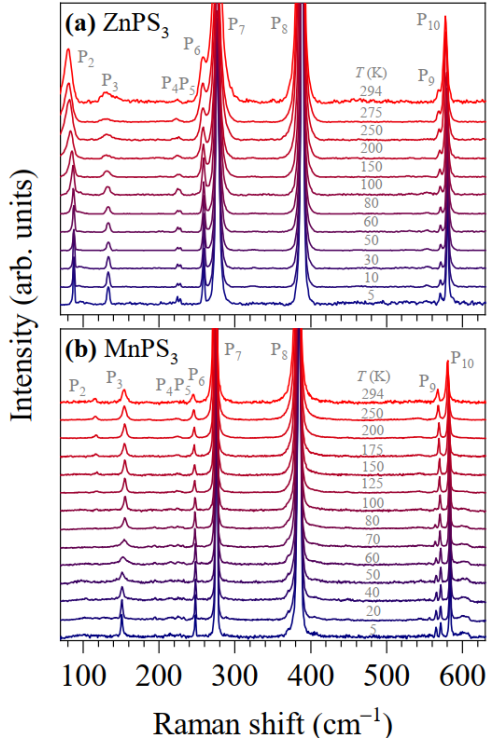

**Fig. 3.**
Raman spectra of ZnPS$_3$ (top panel) and MnPS$_3$ (bottom panel) acquired at different temperatures, from ambient temperature (top spectrum) down to liquid helium temperature (bottom spectrum). First-order Raman A$_g$ and B$_g$ modes are labeled from P$_2$ up to P$_{10}$. A smooth background has been subtracted and all spectra have been vertically shifted for better clarity.

In order to accurately perform an assignment of all the Raman features of MnPS$_3$, a spectrum integrated with a sufficiently long time was acquired at $T = 3.4$ K, and all weak Raman features were rendered visible. This is shown in Fig. 4, together with extracted experimental peak frequencies (shown as colored ticks below the spectrum) and calculated peak frequencies (shown as colored ticks at the bottommost region of the figure). From the figure, it can be seen that the most prominent peaks, labeled from P$_2$ to P$_{10}$ correspond to A$_g$ and B$_g$ Raman-active modes (experimental frequencies are shown in Fig. 4 as red and blue ticks below the spectrum). Second-order Raman features are observed at 544 and 760 cm$^{-1}$, which correspond to 2A$_{g5}$ (P$_7$) and 2A$_{g6}$ (P$_8$) overtones, respectively, as well as 405, 430, 537 and 622 cm$^{-1}$ which we assign to the combination of A$_g$ modes of A$_{g2}$+A$_{g4}$ (P$_3$+P$_6$), A$_{g3}$+A$_{g5}$ (P$_3$+P$_7$), A$_{g4}$+A$_{g5}$ (P$_6$+P$_7$) and A$_{g4}$+A$_{g6}$ (P$_6$+P$_8$), respectively (all second-order Raman peak frequencies are included as black ticks below the spectrum).

Our assignment of second-order modes is in perfect agreement with that previously reported by Peschanskii et al.[40] who proposed that all low-temperature weak Raman features correspond to second-order Raman processes. However, we support that the assignment of other weak features cannot be attributed to a second-order process since either those modes are not observable above $T_N$ or their intensity is comparable to that of second-order modes arising from the combination of the very intense P$_7$ and P$_8$ peaks. For instance, the peak at 450 cm$^{-1}$ previously attributed to a 2A$_{g3}$ overtone is observable at ambient temperature but its intensity is in the same order of magnitude as A$_{g3}$, hence 2 orders of magnitude larger than expected for an overtone (our assigned second-order modes exhibit intensities 2 orders of magnitude lower than their corresponding first-order modes, as typically expected for non-resonant processes). We assigned such a feature to a disorder-activated A$_u$ mode since our calculations predict a very large and narrow density-of-states of this mode (the very flat dispersion of the corresponding phononic branch can be seen in Fig. 2-a) at a frequency of 436 cm$^{-1}$, which is within the −5% range of typical frequency underestimation in DFT calculations within presently used functionals. We assign the remaining peaks to folded phonon modes arising from the V point of the BZ (marked with green ticks below the spectrum). As can be seen in the figure, most of the calculated frequencies at V (shown as green ticks at the bottommost of the figure) agree well with the experimental observations. Remarkably, our calculations predicted a zone-folded mode up to 35 cm$^{-1}$ above P$_{10}$, in good agreement with an experimentally observed band in the range of 14-27 cm$^{-1}$ above P$_{10}$. Moreover, most of the calculated zone-folded modes exhibit frequencies in the 100-290 cm$^{-1}$ range in very good agreement with the experimental observation.

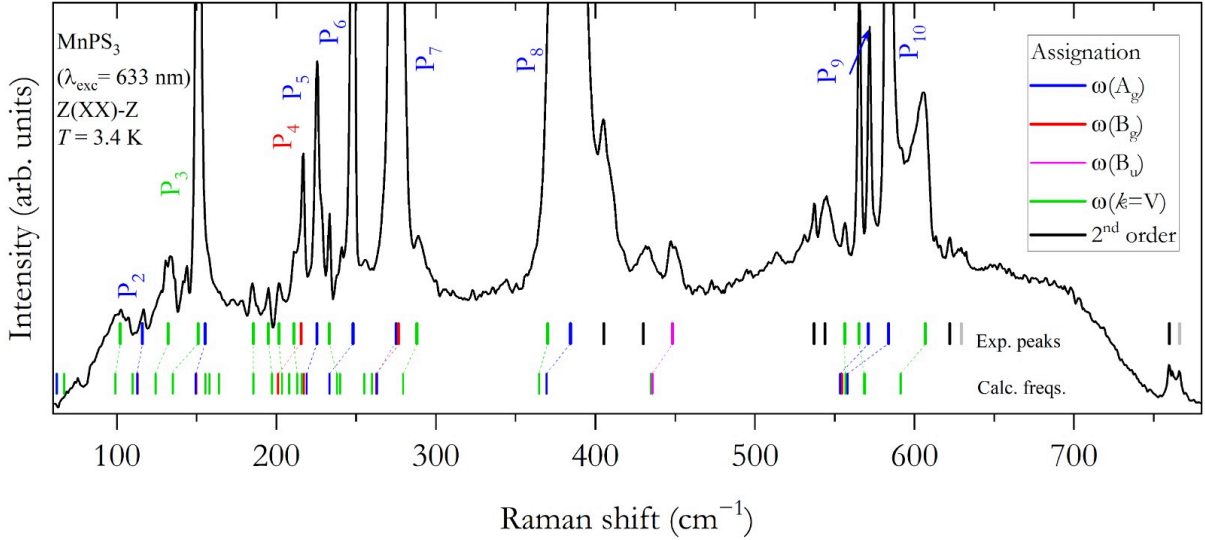

**Fig. 4.** Raman spectrum of MnPS$_3$ acquired at a temperature of 3.4 K. Raman A$_g$ peaks are labeled from P$_2$ to P$_{10}$ in blue, while P$_3$ (at low temperature) is assigned to be a purely folded mode and P$_4$ (labeled in red) is a pure B$_g$ mode which is still visible despite it is symmetry-forbidden for the Z(XX)-Z scattering configuration here used, in Porto's notation. The peak positions of all prominent Raman features are shown with ticks below the spectrum, and their color corresponds to our assignment (dashed lines connect experimental with calculated, at the bottom-most, frequencies). Most peaks have been assigned to either A$_g$, B$_g$, an activated A$_u$ mode, zone-folded modes or second-order Raman modes in blue, red, magenta, green and black ticks, respectively. Only two weak peaks have not been assigned (grey ticks). At the bottommost of the figure calculated frequencies for A$_g$, B$_g$ and modes at the edge of the Brillouin zone (V point) are shown as blue, red and green ticks, respectively.

The zone-folding phenomenon due to antiferromagnetic magnetic ordering is expected to be stronger for phonons involving metallic ion displacements. In this regard, we propose that the vanishing of P$_2$ right below $T_N$ is due to the antiferromagnetic zone-folding mechanism, as well as an abrupt decrease of the frequency of $P_3$ ($\Delta\omega \approx -3$ cm$^{-1}$ as can be seen in Fig. SM3 of the SM). Most of the features (i.e. 13) could be assigned to folded-phonon modes (green ticks in Fig. 4) while 2 high-frequency Raman features remained unassigned (grey ticks in Fig. 4).

***Temperature-dependent Raman measurements of $Mn_xZn_{1-x}PS_3$***

The Raman spectra of $Mn_xZn_{1-x}PS_3$ are shown in Fig. 5 for the whole compositional range at ambient temperature (top panel) and liquid helium temperature (bottom panel). From the figure, it can be seen that the number, frequency and intensity of the peaks depend on the particular compositional content. The splitting of peaks $P_2$, $P_3$ and $P_6$ can be easily observed at both ambient and low temperature, especially for compositions around 50%. This corresponds to an expected two-mode behavior consequence of a flat dispersion of the phononic branches and a somewhat large frequency difference of the modes between the compositional end members (phonon dispersion curves of $ZnPS_3$ and $MnPS_3$ are shown in Fig. 1 and 3, respectively). More interesting is the fact that very weak Raman features show up at ambient temperature for intermediate compositions (maximum signal is found at $x = 50\%$) which are very similar to those of $MnPS_3$ at low temperature (marked with asterisks in Fig. 5). These features cannot be assigned to second order processes owing to their particular frequencies and intensities not matching any two-mode combination or overtone. In the present work, we propose that these features are alloy-activated zone-folded modes in the in-plane direction. Indeed, the alloyed sample $Mn_{0.50}Zn_{0.50}PS_3$ can be regarded, as a first approximation, to the perfectly-ordered mixed-crystal (i.e. $MnZnP_2S_6$) since at low temperatures mean free paths (MFP) of coherent phonons are typically a few orders of magnitude larger than the lattice unit cell. This hypothesis is supported by the fact that the weak Raman features of the alloy exhibit almost identical frequencies and intensities to those of the low-temperature antiferromagnetically-ordered MnPS$_3$. Taking, for instance, Mn$_{0.50}$Zn$_{0.50}$PS$_3$, the zone-folded features are, similarly to MnPS$_3$ at low temperature; *i)* a strong reduction of the P2 peak, *ii)* a peak at 195 cm$^{-1}$, *iii)* a low-frequency tail of P8, *iv)* a band around 600 cm$^{-1}$, as can be seen in Fig. 5 at either ambient or low temperatures (see Fig. SM5 of the SM for an enlarged view). The only substantial difference arises from the fact that a peak 6 cm$^{-1}$ below P$_9$ visible for MnPS$_3$ at low temperature is not visible for Mn$_{0.50}$Zn$_{0.50}$PS$_3$ at neither ambient nor low temperature. This is indeed in agreement with our calculations which predict that this particular folded mode is only present for perfectly ordered structures (s.g. P2/m).

For the particular case of compositions around [x]=50%, the broadening of the Raman peaks is strikingly similar to that of the pure compositional end members, as can be seen in Fig. 5 (for the most intense peak, P$_8$, the full width at half maximum is FWHM = 5 cm$^{-1}$ for all compositions at ambient temperature). This is due to the fact that, unlike conventional bulk crystals where alloying typically increases the peak broadening by around one order of magnitude,[41,42] in alloyed layered compounds structural defects are comparatively lower resulting in a typical peak broadening of a factor of 2 for intermediate compositions.[43] For the particular case of our Mn$_x$Zn$_{1-x}$PS$_3$ samples a small broadening factor of 1.05 at 4K and 1.07 at ambient temperature suggests that the crystallinity is exceptionally good due to the similar size of the Mn and Zn ions which result in each metal ion being properly contained within the octahedral S$_6$ cage.

In order to shed new light on the origin of the structural anomalies of ZnPS$_3$ as well as their impact on the alloy, systematic measurements have been performed on all alloyed samples for temperatures ranging from 4K up to ambient temperature. From Fig. 5-a, it can be seen that Zn-rich samples exhibit broad low-frequency peaks (from $P_2$ to $P_7$), which correspond to vibrations involving mostly the metal cations (for $P_7$, FWHM = 5 cm$^{-1}$ of samples with [Zn]>65% while FWHM = 2.6 cm$^{-1}$ for pure MnPS$_3$ at low temperature). On the other hand, the FWHM of the high-frequency peaks, corresponding to vibrations of the P$_2$S$_6$ units, remains similar at ambient temperature to that of MnPS$_3$. The temperature and compositional dependence of the broadening of the low-frequency $P_7$ and high-frequency peak $P_8$ are shown in the supplementary material (Fig. SM7 and SM8). From these figures, it can be concluded that the structural distortion in ZnPS$_3$ becomes strong at temperatures higher than 100 K and linearly decreases for samples with [Mn] > 35%. Similar non-linear structural effects have been previously reported for perovskite alloys.[44,45]

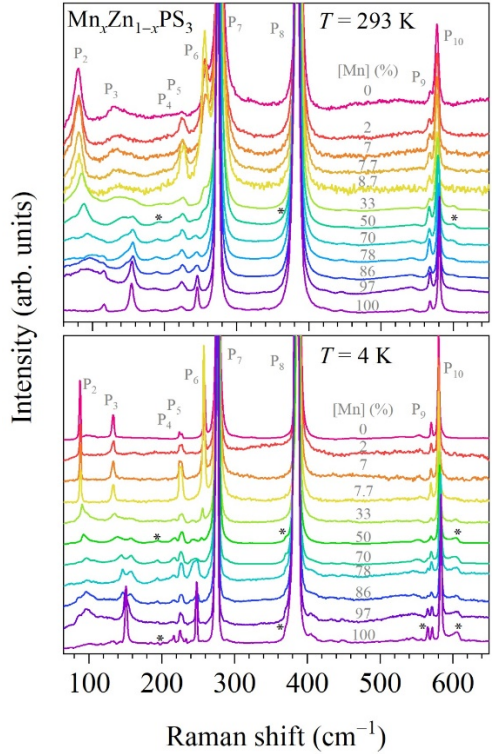

**Fig. 5.**

Raman spectra of different samples of $Mn_xZn_{1-x}PS_3$ covering the whole compositional range acquired at **(a)** ambient temperature and **(b)** liquid helium temperature. All first-order Raman peaks have been labeled from $P_2$ up to $P_{10}$, these correspond to $A_g$ and/or $B_g$ Raman modes since all spectra have been acquired under unpolarized conditions. Weak Raman features showing up at intermediate composition, and below $T_N$ for Mn-rich $Mn_xZn_{1-x}PS_3$ have been marked with asterisks and attributed to zone-folded phonon modes arising from different mechanisms.

For the case of Mn-rich samples, magnetic ordering plays a major role in the lattice dynamics at low temperatures. For instance, the FWHM of the high-frequency $P_8$ peak of $MnPS_3$ remains broad and constant, around 4 cm$^{-1}$, below ≈120 K, (see Fig. SM7-b). We tentatively attribute such a feature to the entrance of a frustrated spin glass phase above the Néel temperature characterized by a short-range spin-spin correlation below 120 K that coincides with a maximum in the susceptibility (as discussed below, on Fig. 8-a),[7,46] in agreement with neutron scattering measurements which revealed scattering from correlations shorter than the scale of the Bragg peaks at 100 K and were attributed to two-dimensional critical fluctuations.[47] Previous works have shown that magnetic ordering strongly affects the frequency and intensity of peak $P_3$ around $T_N$.[40,48] Hence, $P_3$ can be used as a signature to estimate the Néel temperature from Raman measurements alone.

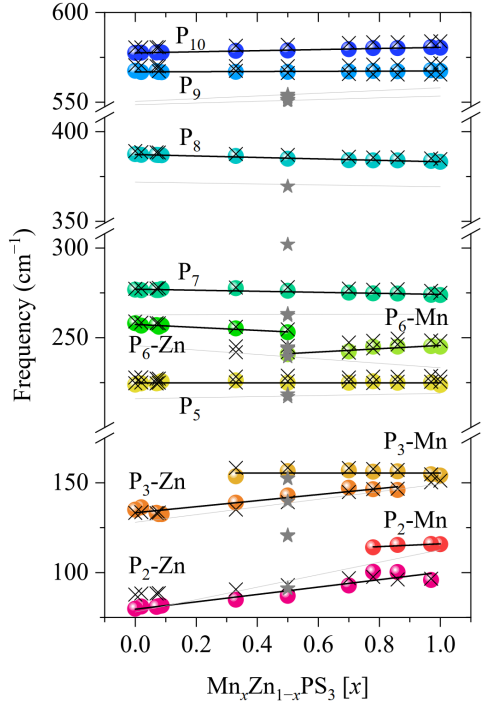

**Fig. 6.**
Compositional dependence of the fitted frequencies of the Raman peaks as measured at ambient temperature (colored symbols) and 4 K (crosses). Linear fits are included for the data acquired at room temperature and peak labels are included accordingly. Calculated frequencies for the compositional end members are shown with thin grey lines. Calculated frequencies for the exact composition of 0.5 are shown as grey stars. Extracted frequencies for the compositional end members are shown in Table 1.

The peak frequencies of spectra in Fig. 5 have been extracted at room and liquid helium temperatures by fitting Lorentzian peaks and are plotted as a function of composition in Fig. 6. As can be seen in the figure, alloy peaks $P_5$, $P_7$-$P_{10}$ exhibit a linear dependency between the compositional end members, while peaks $P_2$, $P_3$ and $P_6$ exhibit a two-phonon mode behavior. For the latter case, their frequencies interpolate from the frequency of a compositional end member to the corresponding local vibrational mode (LVM). In some compositional ranges (mostly for [Mn] > 30%) these peaks coexist and can be observed in the Raman spectra as doublets. From the figure, it can be seen that the frequencies of the peaks at low temperatures (marked as black crosses) are similar to those obtained at ambient temperature (marked as colored symbols), and their frequencies are slightly increased as expected from the intrinsic thermal expansion. However, the frequency

P$_3$ of MnPS$_3$ appears to strongly decrease at low temperature, from an ambient frequency of 156.0 cm$^{-1}$ down to 151.2 cm$^{-1}$, which is a very similar frequency to that corresponding to the extrapolated P$_3$-Zn to [Mn] = 100% (i.e. the frequency of the P$_3$ Zn LVM is 150.2 cm$^{-1}$). Here we sustain that the apparent peak shift is mostly due to a vanishing of a P$_3$-Mn peak, which is not allowed in the AFM ordering, simultaneously accompanied by the re-entrance of a closely-lying P$_3$-Zn peak which is an alloy-like mode. This is clearer for P$_2$ due to the larger separation between the frequency of the LVM (100.2 cm$^{-1}$ at ambient temperature, [$x$]=1) and Mn-P$_2$ (116.0 cm$^{-1}$ at ambient temperature, [$x$]=1). The Mn-P$_2$ peak is only active in the paramagnetic C2/m structure, and vanishes at both; *i*) low temperatures due to AFM ordering and *ii*) ambient temperature with significant alloying (i.e. it vanishes for alloy concentrations [$x$]<0.75).

The peak frequencies shown in Fig. 6 have been linearly fitted for measurements performed at ambient and low temperatures (i.e. 4 K). Extracted experimental frequencies are provided in Table 1 together with the calculated figures. For comparison purposes, calculated frequencies of the A$_g$ modes of MnPS$_3$ and ZnPS$_3$ have been linearly interpolated and plotted in Fig. 6 as thin grey lines. From the figure, it can be seen that our calculations underestimate the frequency values by up to −5%, especially those of modes involving the movement of light ions. A ~5% underestimation of the optical phonon frequencies is well-documented for the PBEsol functional applied to two-dimensional compounds.[49] Fig. 6 includes the calculated frequencies of the modes of perfectly mixed crystal MnZnP$_2$S$_6$ (a total of 15+14=29 optical modes with A and B symmetries, respectively, are shown as grey star symbols in the figure), where each metal is exchanged to first neighbors in-plane (the corresponding space group is C2). Calculations predict that $P_5$, $P_7$, $P_8$, and $P_9$ exhibit a one-mode behavior and with a frequency that almost linearly interpolates between the compositional end members while $P_2$, $P_3$ and $P_6$ exhibit a two-mode behavior (the one- and two-mode behavior can be seen from the calculated frequencies for $[x]=0.5$, shown as grey stars in Fig. 6), in excellent agreement with experiments.

|  |  | MnPS$_3$ | | | ZnPS$_3$ | | |
| --- | --- | --- | --- | --- | --- | --- | --- |
| Peak label | Phonon mode | Experiment $T = 293$ K | Experiment $T = 10$ K | Calculations | Experiment $T = 293$ K | Experiment $T = 10$ K | Calculations |
| P$_1$ | B$_{g1}$ | – | – | 62.3 | – | – | 53.7 |
| P$_2$-Zn | A$_{g1}$ B$_{g2}$ | – | – | – | **79.6** | **87.9** | 78.0 79.0 |
| P$_2$-Mn | A$_{g1}$ B$_{g2}$ | **116.0** | – | 112.9 112.9 | – | – | – |
| P$_3$-Zn | A$_{g2}$ B$_{g3}$ | – | – | – | **133.4** | **132.0** | 128.0 130.6 |
| P$_3$-Mn | A$_{g2}$ B$_{g3}$ | **155.5** | **156.0** | 149.5 150.1 | – | – | – |
| P$_4$ | B$_{g4}$ | – | – | 201.3 | – | – | 185.8 |
| P$_5$ | A$_{g3}$ B$_{g5}$ | **224.9** | **225.5** | 219.3 217.2 | **224.9** | **224.8** | 216.1 214.8 |
| P$_6$-Mn | A$_{g4}$ | **245.7** | **249.0** | 233.0 | – | – | – |
| P$_6$-Zn | A$_{g4}$ | – | – | – | **257.4** | **258.8** | 245.7 |
| P$_7$ | A$_{g5}$ B$_{g6}$ | **274.2** | **276.4** | 263.1 262.4 | **277.1** | **277.8** | 262.8 261.8 |
| P$_8$ | A$_{g6}$ | **383.2** | **384.9** | 369.3 | **387.2** | **388.8** | 372.0 |
| P$_9$-Mn | A$_{g7}$ B$_{g7}$ | **567.4** | **571.2** | 553.4 554.6 | – | – | – |
| P$_9$-Zn | A$_{g7}$ B$_{g7}$ | – | – | – | 566.9 | 570.0 | 548.6 551.3 |
| P$_{10}$ | A$_{g8}$ | **580.6** | **584.0** | 558.0 | **577.5** | **580.4** | 550.5 |

**Table 1.** Experimental and calculated phonon frequencies (in units of cm$^{-1}$) of MnPS$_3$ (left) and ZnPS$_3$ (right). Experimental values are obtained from a linear fit of the compositional dependence of the Raman features for alloyed samples of all compositions. Experimental figures in bold correspond to peaks measured for the compositional end members.

***Determination of the Néel temperature from Raman spectroscopy***

In order to determine the Néel temperature in the alloy several Raman features can be used, such as the broadening of low-frequency peaks, peak positions, the appearance and disappearance of new phonon modes due to the antiferromagnetic zone-folding phenomenon, or variations in relative intensities. However, we found, in agreement with previous works,[22,40,48] that the most temperature-sensitive signature arises from the frequency of the $P_3$-Mn mode (a detailed illustration of the evolution of $P_3$ with temperature for $MnPS_3$ is shown in Fig. SM3 of the SM). The physical origin of the $P_3$ mode remains a matter of debate despite the two-magnon hypothesis (Raman-active magnons have been observed in similar compounds such as $FePS_3$[50]) has been ruled out due to its discrepancy in energy (i.e. 177 cm$^{-1}$ for a two-magnon as measured by neutron diffraction studies,[51] far from the experimental 154 cm$^{-1}$ value of $P_3$). In this regard, a few alternative interpretations have been presented, including; *i)* the presence of a more complex magnetic sublattice lacking magnetism for some magnetic ions,[40] *ii)* a two-particle phonon-magnon excitation,[40] *iii)* a spin-related magnetic-antiferromagnetic phase transition resulting in the disappearance of the $P_3$ mode and a re-entrance of a second mode located 4 cm$^{-1}$ below $P_3$, that shows up below a temperature of 50 K, close to the critical temperature of 55 K where $MnPS_3$ might exhibit Heisenberg XY-like ordering,[48] *iv)* a single mode that broadens and redshift as a consequence of spin-phonon coupling where fluctuations are responsible for disrupting the coherence of the lattice vibrational modes and shortening the lifetimes, resulting in a sharp increase of the linewidth,[22] and *v)* a Fano resonance between a phonon peak and a two-magnon continuum.[52] Since Mn-rich samples exhibit a $P_3$ doublet (see Fig. 5-b) with relative intensities strongly dependent on the temperature, it is clear to us that $P_3$ is indeed two closely-lying modes with different physical origins. Hence, the apparent broadening and shift towards lower frequencies of the $P_3$ mode in pure $MnPS_3$ are mostly explained by the increase in the intensity of the low-frequency peak. Since the energy separation between both peaks is around 3 cm$^{-1}$ and their broadening is around 4 cm$^{-1}$ for pure $MnPS_3$ and as large as 10 cm$^{-1}$ for a sample with [Mn] = 78%, the $P_3$ doublet in $MnPS_3$ is not possible to be experimentally resolved. However, while the low-frequency peak corresponds to a folded mode (with increasing intensity with reduction of temperature, as also observed for the alloy in Fig. 7-a), the high-frequency $P_3$ peak is still susceptible to hybridize with the two-magnon continuum. Indeed, the frequency of the peak $P_3$-Mn decrease below $T_N$ for the alloy. Here we propose that the zone-folding phenomenon takes place for both, $MnPS_3$ and $Mn_xZn_{1-x}PS_3$ by two physically distinct mechanisms. From one side, the antiferromagnetic arrangement of NNs and from the other, the intercalation of different atomic species in the alloy. Both mechanisms result in almost identical Raman signatures and add to each other. This can be seen for sample $Mn_{0.78}Zn_{0.22}PS_3$ in Fig. 7-a; *i)* the ambient-temperature damped $P_2$ peak fully vanishes at lower temperatures, *ii)* the relative intensity of $P_3$-Zn with respect $P_3$-Mn increase with decreasing temperature and *iii)* the weak band around 600 cm$^{-1}$ increases its intensity with decreasing temperature.

Following the previous argumentation, it is possible to determine $T_N$ from variations in relative intensities between the $P_3$ doublet, or small variations of peak frequency. The temperature-dependency of the frequency of the Mn-like $P_3$ peak is plotted in Fig. 7-b for samples with [Mn] content larger than 78%. As can be seen in the figure, the frequencies increase with decreasing temperature, as expected as a consequence of thermal lattice contraction, but at a certain temperature, an abrupt decrease in frequency (and intensity) takes place, especially for the pure $MnPS_3$ compound. From this analysis it can be seen that the Néel temperature decrease with decreasing [Mn] content. This is expected since short-range magnetic interaction is quenched in the diamagnetic lattice. From these results it seems clear that the Néel temperature of [Mn]>97% is around 78 K and for 78%<[Mn]<86%, $T_N$ is 49 K. These results are in excellent agreement with our SQUID measurements; $T_N$ = 48.5 K for a sample with 78% Mn content. The compositional dependence of $T_N$ is plotted in panel c of Fig. 7. From this figure it can be seen that excellent agreement is reached between previously[15] and presently reported SQUID measurements, measurements using a Faraday balance[14] and our Raman measurements. By linearly interpolating the phase diagram it seems that no long-range magnetic ordering would take place for samples with [Mn]<30%.

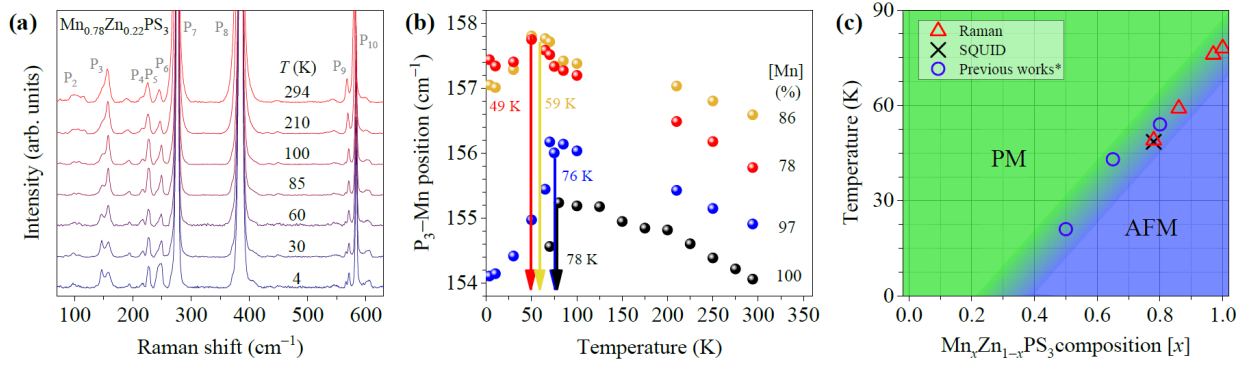

**Fig. 7. (a)** Raman spectra of a $Mn_{0.78}Zn_{0.22}PS_3$ sample acquired at different temperatures, from 4 K (bottom) up to ambient temperature (top). **(b)** Dependence on temperature of the fitted frequency of the mode $P_3$-Mn. **(c)** Magnetic phase diagram including experimental data obtained in the present work by means of Raman spectroscopy (triangles), SQUID measurements (cross) and *magnetic measurements performed by Goossens et al.[15] and Chandrasekharan et al.[14] (circles).

### SQUID magnetic-field measurements

Magnetic susceptibilities of MnPS$_3$ and high alloying ratios of Mn$_x$Zn$_{1-x}$PS$_3$ ($0<x\leq1$) were measured over a wide temperature range by a SQUID magnetometer. Fig. 8-a reveals the temperature dependences of mass magnetic susceptibilities χ of MnPS$_3$ bulk single crystal in two directions; in-plane (χ$_{//ab}$) and out-of-plane (χ$_{\perp ab}$) at a constant magnetic field (100 Oe). For Bulk MnPS$_3$, each Mn$^{2+}$ ion is coupled antiferromagnetically with its nearest neighbor in the two-dimensional plane

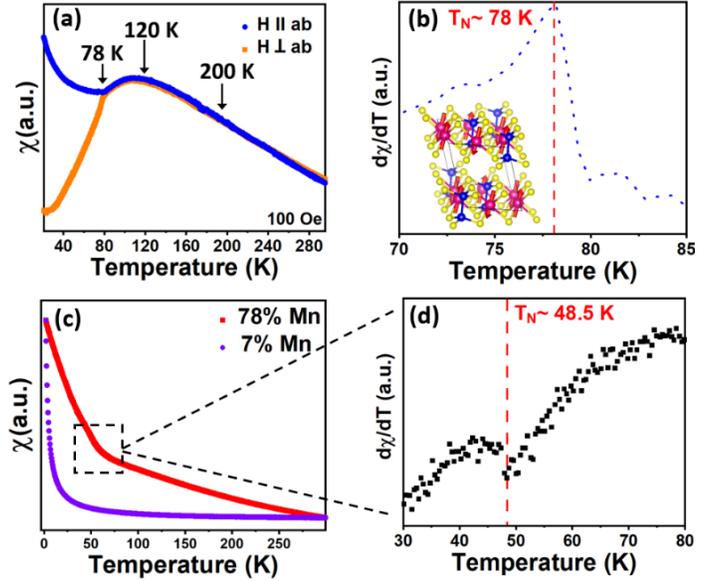

**Fig. 8. Identification of the Néel transition temperature in bulk Mn$_x$Zn$_{1-x}$PS$_3$ crystals.** (a) Temperature dependence of the magnetic susceptibility (χ) of bulk MnPS$_3$ crystals with the magnetic field applied in-plane (blue curve) and perpendicular to the plane (orange curve). The black arrow points to the temperature below which the two curves start to deviate, which allows the Néel temperature ($T_N$) identification. (b) dχ/dT of the in-plane measurement in (a) with an inflection point at $T_N \approx 78$ K. The inset illustrates the antiferromagnetic spins arrangement below $T_N$. (c) Temperature dependence of the magnetic susceptibility of Mn$_{0.78}$P$_{0.22}$S$_3$ (red) and Mn$_{0.07}$Zn$_{0.93}$PS$_3$ (purple). The zoom area is focused on the inflection point at low temperatures. (d) dχ/dT of the in-plane measurement of Mn$_{0.78}$Zn$_{0.22}$PS$_3$ in (c) with an inflection point at $T_N \approx 48.5$ K.

(in-plane). The coupling between adjacent planes is ferromagnetic (out-plane). MnPS$_3$ antiferromagnetic system exhibits a characteristic temperature, termed Néel temperature ($T_N$), at which the long-range magnetic order is finally lifted. $T_N \sim 78$ K is determined from the discontinuity in the first derivative $(\partial\chi/\partial T)_H$ shown in Fig. 8-b. The antiferromagnetic arrangement below $T_N$ is illustrated in the inset. For MnPS$_3$, the susceptibility dependence on strength of the magnetic field can be divided into three different regions: (i) below $T_N$~78 K, $\chi_{//}$ tends to zero with decreasing temperature, while $\chi_\perp$ increases gradually. The anisotropic antiferromagnetic order below $T_N$ can be originated from: First, the range of exchange interactions is beyond the nearest neighbor, arising from a non-negligible spin-orbit coupling. Second, the dipolar anisotropy dictates the magnetization axis and provides the system with a weak Ising character.[53] (ii) Above $T_N$; from 78 to 120 K, the susceptibilities behave equally, where an isotropic and broad hump is observed at 120 K (T ~ 3/2 $T_N$). The broad maximum at 120 K can be explained by the change in magnetic structure, reflecting a small single-ion anisotropy order due to short-range spin−spin correlation in the ab plane,[7,46,47] which is typical for low dimensional 2D magnetic systems. (iii) From 120 to 200 K and to ambient temperature, a linear behavior is noted, demonstrating a paramagnetic phase order in MnPS$_3$.

Fig. 8-c displays the temperature dependence of the magnetic susceptibility of highly Mn-alloyed Mn$_{0.78}$Zn$_{0.22}$PS$_3$ ($x$ = 0.78) and Mn-diluted Mn$_{0.07}$P$_{0.93}$S$_3$ ($x$ = 0.07) crystals with the magnetic field applied in-plane at constant magnetic fields of 1000 Oe and 100 Oe, respectively. For Mn$_{0.78}$P$_{0.22}$S$_3$ (red curve), the paramagnetic region is extended toward a lower temperature compared to MnPS$_3$ and $T_N$ is shifted. $T_N$ = 48.5 K was estimated from the discontinuity in the first derivative ($\partial\chi/\partial T$) shown in Fig. 8-d. It is worth noticing that the hump in the susceptibility at $T>T_N$ vanishes, which could be explained by lowering the Mn nearest-neighbor interactions in the Mn$_{0.78}$Zn$_{0.22}$PS$_3$ system leading to the breakdown of the short-range spin-spin correlation.[15] Below $T_N$, the susceptibility increases due to long-range exchange interaction. For Mn-diluted Mn$_{0.07}$Zn$_{0.93}$PS$_3$ crystal, no long-range

antiferromagnetic transition was observed (purple curve), as in this case a significant number of the spins no longer belong to the infinite cluster so the magnetic susceptibility matches the behavior of a weak paramagnetic system.

## III CONCLUSIONS

We provided an assignment of the Raman-active phonon modes of $Mn_xZn_{1-x}PS_3$ and accurately described its compositional dependence in terms of one- and two-mode phonon behavior. With the aid of first-principle calculations, all Raman features have been assigned either to first-order modes, second-order modes, a silent mode, or zone-folded modes activated through different physical mechanisms. These are either the mixing of two different metallic chemical species or inducing antiferromagnetic ordering of the metal ions at low temperatures, both resulting in the reduction of crystal symmetry and similar Raman features. Low-temperature Raman and magnetic measurements allowed to determine the Néel temperatures of Mn-rich $Mn_xZn_{1-x}PS_3$ samples and complete the corresponding phase diagram. Hence, we confirm that Raman spectroscopy is a valid tool to identify magnetic transitions in two-dimensional alloyed systems and relevant Raman signatures have been identified for $Mn_xZn_{1-x}PS_3$. Finally, we report abnormally broad low-frequency Raman peaks for $ZnPS_3$ at ambient temperature, which is consistent with previous claims on the presence of a second-order Jahn-Teller structural distortion.

## IV. METHODS

*Preparation and structural characterization*

Single crystalline $Mn_xZn_{1-x}PS_3$ (0≤$x$≤1) samples were grown via vapor transport synthesis (VTS) without any transporting agent.[15,54] Selected amounts of powder elements (Zn, Mn, P and S) were calculated to obtain around 1g of substrate mixture. The mixture was grounded in an agate mortar and sealed in an evacuated quartz ampoule at a pressure below 3.5×10⁻⁵ Torr. The reaction duration was set for one week, and took place in a two-zone furnace with a gradient of temperature where the substrate zone was kept at 650 °C and the cold deposition zone was at 600 °C. For pure $ZnPS_3$ crystals, lower temperatures were used, which are 600 °C for the substrate zone and 550 °C for the deposition zone. Crystals from the deposition zone were collected for further characterization. The composition of every crystal used in this research was investigated by EDX as the material from the deposition zone (i.e. recrystallized $Mn_xZn_{1-x}PS_3$) may have a different [Mn]/[Zn] ratio compared to the initially used substrate zone mixture. The $Mn_xZn_{1-x}PS_3$ compounds crystallized into a bulk monoclinic layered structure with a space group of C2/m. Fig. 9-c shows a side-view (top panel) and top-view (bottom panel) of the monoclinic crystal system. Single crystal X-ray diffraction (XRD) measurements allowed to extract the lattice parameters; $a$ = 6.0780 Å, $b$ = 10.5332 Å, $c$ = 6.7887 Å, β = 107.122° for $MnPS_3$ and $a$ = 5.9576 Å, $b$ = 10.3252 Å, $c$ = 6.7648 Å, β = 107.182° for $ZnPS_3$, in agreement with previous reports.[32,33] Abnormally high atomic displacement parameters (ADP) were found for the Zn atoms in $ZnPS_3$ ($U_{eq}$ = 0.030(1) Å) compared to the Mn atoms in $ZnPS_3$ ($U_{eq}$ =

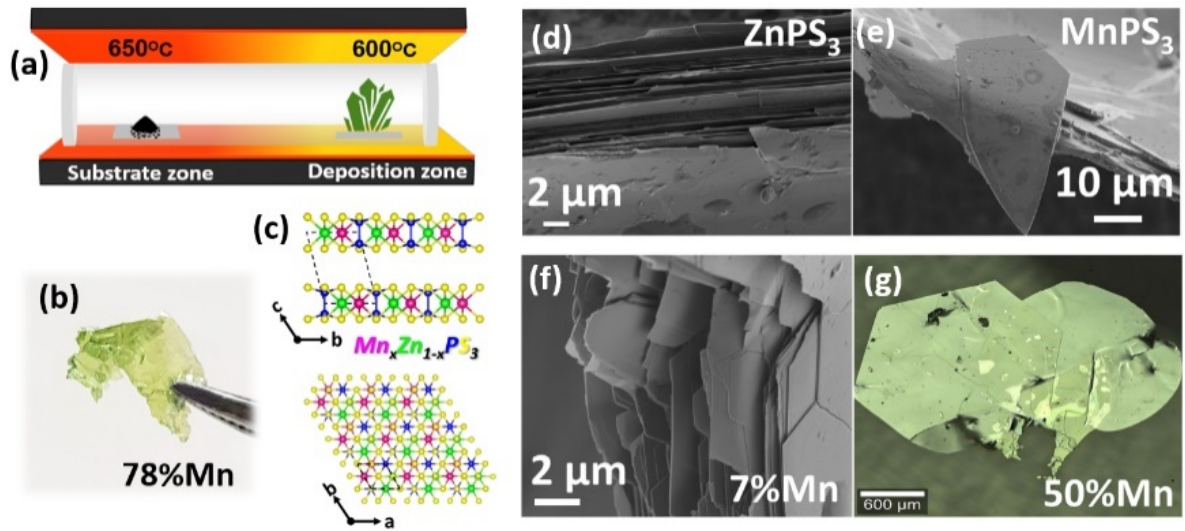

**Fig. 9. Preparation and characterization of $Mn_xZn_{1-x}PS_3$ crystals.** (a) Preparation method of $Mn_xZn_{1-x}PS_3$ crystals via vapor transport synthesis using a two-zone furnace. (b) A photograph $Mn_{0.78}Zn_{0.22}PS_3$ crystal seized from the deposition zone. (c) Side-view and (top panel) top-view (bottom panel) of the monoclinic crystal system which belongs to the $Mn_xZn_{1-x}PS_3$. (d-f) HR-SEM images of $ZnPS_3$, $MnPS_3$ and $Mn_{0.07}Zn_{0.93}PS_3$ bulk crystals. (g) Magnified optical microscope image of $Mn_{0.5}Zn_{0.5}PS_3$ flakes.

0.01359(8) Å). These values are in good agreement with those previously reported by X-ray diffraction measurements.[37]

**Composition determination by EDX and SEM morphology characterization**

High-resolution scanning electron microscope (HR-SEM) images were registered with Zeiss Ultra-Plus FEG-SEM. Energy-dispersive X-ray spectroscopy (EDX) spectra were acquired with FEI E-SEM Quanta 200. Both measurements were obtained at accelerating voltage of 20 kV. EDX spectra were collected to quantitatively analyze the composition and to estimate the [Mn]/[Zn] atomic ratios of the metal in alloyed $Mn_xZn_{1-x}PS_3$ samples.

The morphology of bulk $Mn_xZn_{1-x}PS_3$ was examined using HR-SEM and an optical microscope. A selected photograph of a sample with [$x$] = 78% is shown in Fig. 9-b and represents a greenish-colored crystal. Fig. 9-d,e,f show HR-SEM micrographs of the $ZnPS_3$, $MnPS_3$, and $Mn_{0.78}Zn_{0.22}PS_3$ crystals, respectively, highlighting the layered nature of the pure and alloyed compounds. Cleavage angles of 30° and 60° can be observed, with the longest sides typically corresponding to the *b* axis. Fig. 9-g shows an optical microscope photograph of $Mn_{0.50}Zn_{0.50}PS_3$ flakes, presenting their semi-hexagonal shape and large surface area.

## *Superconducting quantum interference device*

Superconducting quantum interference device (SQUID) measurements were performed in the Quantum Matter Research (QMR) center in the Technion using the SQUID magnetometer Quantum Design MPMS3 which provides a sensitivity of less than $5 \times 10^{-8}$ emu. Dc magnetic susceptibilities were measured under an external applied field of 100 or 1000 Oe, from 1.8 K to 380 K.

## *Raman measurements.*

Raman measurements have been performed using a micro-Raman spectrometer coupled to the inVia™ Reflex Raman Microscope system from Renishaw. All Raman measurements shown in the main manuscript have been acquired in the backscattering configuration. Polarized Raman measurements have been performed under the parallel and crossed backscattering configurations as well as using different excitation energies. The 632.8-nm excitation line of a He-Ne laser was selected for all reported experiments in the main manuscript due to its enhanced Raman signal and improved spectral resolution. Since selection rules in TMPTs are not strictly respected (see Fig. SM4) all measurements presented in the main manuscript were acquired unpolarized configuration (except otherwise indicated in the figure). Low-temperature measurements have been performed using an Oxford Microstat-Hire open-cycle He cryostat together with an Oxford Mercury ITC controller. A THMS600 Linkam stage was used for the high-temperature measurements.

**Lattice-dynamical calculations**

Unit cell optimization and relaxation of atomic positions were performed by calculations based on the density functional theory (DFT) using the Vienna ab-initio Simulation Package (VASP) which employs a projector-augmented wave (PAW) basis set.[55,56] The exchange-correlation functional revised for solids PBEsol[57] was used with the energy cutoff for plane-wave expansion set to 500eV. Following previous DFT calculations on this compounds(10,58,59), for all the calculations, except those relative to $ZnPS_3$, the DFT+$U_{eff}$ rotational invariant approach of Dudarev[58] was used to characterize the on-site Coulomb repulsion between the 3d electrons of the Mn atoms ($U_{eff}$ = 5eV). For $MnPS_3$ a Néel-type antiferromagnetic order was assumed and a convergence criterion of $5.0 \times 10^{-4}$ eV/Å for the forces was used in all the relaxations. Calculations of the $ZnPS_3$ and $MnPS_3$ pure compounds were performed using the primitive cell defined as $\mathbf{a_p}=(\mathbf{a}-\mathbf{b})/2$, $\mathbf{b_p}=(\mathbf{a}+\mathbf{b})/2$ and $\mathbf{c_p}=\mathbf{c}$, where $\mathbf{a}$, $\mathbf{b}$ and $\mathbf{c}$ are the lattice vectors of the conventional C2/m cell as provided elsewhere,[25] with a Brillouin zone integration scheme of 12×12×10. For the mixed $Mn_{0.50}Zn_{0.50}PS_3$ alloys the conventional cell was used with a k-mesh of 10×7×10. For the alloys ferromagnetic and antiferromagnetic orderings have been considered. In the antiferromagnetic P2/m P2/c structures the two Mn atoms of the primitive unit cell are assumed to have opposite spins and for the C2 structure the spins of the two Mn atoms related by the centering are opposite. Phonon dispersion branches were calculated using the PHONOPY package[59] applying finite displacements in a supercell of 4×4×4 times the primitive cell and sampling the Brillouin zone only in the Γ point.

**DATA AVAILABILITY**

All data derived from the experiments and calculations of this study are available from the corresponding author upon reasonable request.

**ACKNOWLEDGMENTS**

F.H., E.R. and E.L. acknowledge the contribution of the Quantum Research Matter (QMR) center and Dr. Anna Eyal for performing the SQUID measurements. The authors would like to thank prof. Francisco Javier Zúñiga Lagares for his contribution on X-ray diffraction measurements on the samples here studied. I.E. acknowledges the financial support of the "Ministerio de Ciencia e Innovación" (PID2019-106644GB-I00)" and the Basque Country Government (IT1358-22). A. K. B. and E. L. were supported by the European Commission via the Marie Skłodowska-Curie action Phonsi (H2020-MSCA-ITN-642656).

**AUTHOR CONTRIBUTIONS**

R. O. wrote the paper, performed the low-temperature Raman experiments and data analysis. I. E. carried out first-principles calculations and contributed to the theoretical part. Y. A. provided the facilities and means for sample growth, which was performed by E. R., F. H. and A. K. B. The SQUID magnetic measurements were performed by E. R. and F. H. and contributed to the methods section. M. G. and E. L. planned and coordinated the research. All authors discussed the results and commented on the paper.

# ADDITIONAL INFORMATION

**Competing interests:** The Authors declare no Competing Financial or Non-Financial Interests.